\documentstyle[12pt,epsf,epsfig,wrapfig]{article}
\begin{document}
\begin{center}
{\bfseries TRANSVERSE ASYMMETRY IN HADRON LEPTOPRODUCTION AT SMALL $x$}
\vskip 5mm
S.V.Goloskokov
\vskip 5mm
{\small
{\it
Bogoliubov Laboratory of Theoretical  Physics,
 Joint Institute for Nuclear Research,
Dubna 141980, Moscow region, Russia
}
\\
{\it
E-mail: goloskkv@thsun1.jinr.ru
}}
\end{center}
\vskip 5mm

\begin{abstract}
We  consider  double spin asymmetries for longitudinally
polarized leptons and transversally polarized protons in diffractive vector meson and $Q
\bar Q$ production for the HERMES energy range on the basis of the two-gluon model.
The asymmetry predicted for meson production is quite small. Large asymmetry is expected for $Q
\bar Q$ production.
\end{abstract}
\vskip 8mm

Study of  the hadron structure  is a fundamental problem of modern
physics. One of the important objects here is skewed parton
distributions (SPD) in a nucleon which are the distributions generalized
to the case of non-forward scattering \cite{rad-j}.
At small $x$, these distributions can be investigated in diffractive processes.
In the case of diffractive charm quark production including
$J/\Psi$ reactions, the predominant contribution is determined by
the two-gluon exchange (gluon SPD). Thus diffractive heavy quark production should
play a keystone role in the future study  of the gluon distribution ${\cal F}_x(x)$ at
small $x$. Processes where  light
quarks appear should include the quark distribution functions as well as the gluon one.
 In  future, it will be an excellent possibility to study spin effects with
transversally polarized target at HERMES. Such experiments should shed  light on the
polarized parton distributions responsible for the transverse spin effects in the
hadron.

In this report, we consider  double spin asymmetries for
longitudinally polarized leptons and transversally polarized
protons in diffractive vector meson and $Q \bar Q$ production in the
HERMES energy range. The two-gluon exchange model with the
spin-dependent $gg$-proton coupling (\ref{ver}) will be used.
We discuss the connection of two-gluon model with skewed
gluon distribution.

To study spin effects in the diffractive hadron production, one
should know the structure of the two-gluon coupling with the
proton at small $x$.  We use the form
\begin{eqnarray}\label{ver}
V_{pgg}^{\alpha\beta}(p,t,x_P,l_\perp)&=& B(t,x_P,l_\perp) (\gamma^{\alpha}
p^{\beta} + \gamma^{\beta} p^{\alpha}) +\nonumber\\
&+&\frac{i K(t,x_P,l_\perp)}{2 m} (p^{\alpha} \sigma ^{\beta \gamma} r_{\gamma}
+p^{\beta} \sigma ^{\alpha \gamma} r_{\gamma})
+i D(t,x_P,l_\perp)\epsilon^{\alpha\beta\delta\rho}p_{\delta}\gamma_{\rho}\gamma_{5}.
\end{eqnarray}
Here m is the proton mass, $p$ is the proton momentum, $x_P$ is a part of this momentum carried
by the two-gluon system, and $l_\perp$ is the gluon transverse momentum.
The first two terms of the vertex (\ref{ver}) are symmetric in
$\alpha,\beta$ indices. The structure proportional to $B(t,...)$ in (\ref{ver})
determines the spin-non-flip contribution. The term
$\propto K(t,...)$ leads to the transverse spin-flip at
the  vertex.   The asymmetric structure in
(\ref{ver}) is proportional to $D \gamma_{\rho}\gamma_{5}$ and
can be associated with $\Delta G$. It should contribute to the
double spin longitudinal asymmetry $A_{ll}$.

We study a simple model  for the amplitude of the
$\gamma^\star \to V$ transition.  The virtual photon is
going to the $q \bar q$ state, and the $q \bar q \to V$ amplitude
is described by a nonrelativistic wave function
\cite{rysk}. In this approximation, quarks have the same
momenta  equal to half of the vector meson momentum and mass
$m_q=m_V/2$. A model like that should be applicable for heavy ($J/\Psi$) production.
Gluons from the two-gluon-proton structure (\ref{ver}) are coupled with the single
and different quarks in the quark loop. This ensures the
gauge invariance of the final result.

The spin-average and spin-dependent cross sections of the
vector meson leptoproduction with
longitudinal polarization of a lepton and different transverse polarization of
the proton are determined by  the relation
\begin{equation}\label{ds0}
d \sigma(\pm) =\frac{1}{2} \left( d \sigma({\rightarrow}
{\Uparrow}) \pm  d \sigma({\rightarrow}
{\Downarrow})\right).
\end{equation}
The cross section $d \sigma(\pm)$ can be written in the form
\begin{equation}\label{ds}
\frac{d\sigma^{\pm}}{dQ^2 dy dt}=\frac{|T^{\pm}|^2}{32 (2\pi)^3
 Q^2 s^2 y}.
\end{equation}
 For the spin-average  amplitude square we find
\begin{equation}\label{t+}
|T^{+}|^2=  s^2\, N \,\left( (1+(1-y)^2) m_V^2 + 2(1 -y) Q^2
\right) \left[ |\tilde B|^2+|\tilde K|^2 \frac{|t|}{m^2} \right].
\end{equation}
 Here $N$  is the
normalization factor, the term proportional to $(1+(1-y)^2) m_V^2$ represents
the contribution of the virtual photon with transverse
polarization. The $2(1 -y) Q^2$ term describes the effect of
longitudinal photons. The $\tilde B$  and $\tilde K$ functions are
expressed through the integral over transverse momentum of the
gluon. The function $\tilde B$ is determined by
\begin{eqnarray}\label{bt}
\tilde B&=&\frac{1}{4 \bar Q^2} \int \frac{d^2l_\perp
(l_\perp^2+\vec l_\perp \vec \Delta) B(t,l_\perp^2,x_P,...)}
{(l_\perp^2+\lambda^2)((\vec l_\perp+\vec
\Delta)^2+\lambda^2)[l_\perp^2+\vec l_\perp \vec \Delta
+\bar Q^2]} \nonumber\\
&\sim& \frac{1}{4 \bar Q^4}\int^{l_\perp^2<\bar Q^2}_0
\frac{d^2l_\perp (l_\perp^2+\vec l_\perp \vec \Delta) }
{(l_\perp^2+\lambda^2)((\vec l_\perp+\vec \Delta)^2+\lambda^2)}
B(t,l_\perp^2,x_P,...),
\end{eqnarray}
with $\bar Q^2=(m_J^2+Q^2+|t|)/4$. The term $(l_\perp^2+\vec
l_\perp \vec \Delta)$ appears in the numerator of (\ref{bt})
because of the cancellation between the graphs where gluons are
coupled with the single and different quarks. The $\tilde K$
function is determined by a similar integral.
The integral (\ref{bt}) can be connected with the gluon
SPD as
\begin{equation}\label{fspd}
{ {\cal F}^g_{x_P}(x_P,t)}=\int^{l_\perp^2<\bar Q^2}_0
\frac{d^2l_\perp (l_\perp^2+\vec l_\perp \vec \Delta) }
{(l_\perp^2+\lambda^2)((\vec l_\perp+\vec \Delta)^2+\lambda^2)}
B(t,l_\perp^2,x_P,...) \propto \tilde B.
\end{equation}
We find that $B(l_\perp^2,x_P,...)$ is the nonintegrated spin-
average gluon distribution. The $\tilde K$ function is proportional to
the ${\cal K}^g_{x_P}(x_P,t)$ distribution.
Determination of the gluon distribution functions can be found e.g. in
\cite{kroll-da}

The spin-dependent amplitude square looks like
\begin{equation}\label{t-}
|T^{-}|^2=  s |t| N \frac{\vec Q \vec S_\perp}{4 m}\left(Q^2+ m_V^2+|t| \right)
\left[ \tilde B \tilde K  \right] .
\end{equation}
The $A_{lT}$ asymmetry for vector meson production  is determined by
\begin{equation}
\label{asy} A_{lT}=\frac{\sigma(-)}{\sigma(+)} \sim
\frac{\vec Q \vec S_\perp}{4 m}\; \frac{y x_P |t|}{ (1+(1-y)^2) m_V^2 + 2(1 -y) Q^2} \;
\frac{\tilde B \tilde K}{ |\tilde B|^2+|\tilde K|^2 |t|/m^2}.
\end{equation}

This  asymmetry is approximately proportional to the ratio of $\tilde K/\tilde B
\propto {\cal K}^g_{x_P}(x_P)/{\cal
F}^g_{x_P}(x_P)$ and generally can be used to obtain information on ${\cal K}^g$ SPD
from experiments with transversely polarized target.
The expected value of $\tilde K/\tilde B$ ratio can be estimated
from elastic scattering. It is about .1  in the models
\cite{gol_kr,gol_mod} which predict asymmetries similar to
that observed experimentally \cite{krish-f}. We use $\tilde K/\tilde B=0.1$
in our estimation. Unfortunately, it is found  that the
$A_{lT}$ asymmetry in the diffractive vector meson
production contains the small coefficient $x_P=(Q^2+
m_V^2+|t|)/(y s)$. The asymmetry predicted for $J/\Psi$
production  at
HERMES energies is shown in Fig.1 for the case when
the transverse part of photon momentum is parallel
to the target polarization $S_\perp$. Simple estimations on the basis of
(\ref{asy}) for $\rho$ meson production are shown
there too. Expected mass dependence is quite weak.
However, this result is obtained for the
nonrelativistic meson wave function
which is not so good approximation for light meson production. At
HERA energies, asymmetry will be extremely small.
\medskip

\begin{minipage}{7.3cm}
\phantom{aa}
\vspace{.3cm}\hspace{-.5cm}
\epsfxsize=6.8cm
\centerline{\epsfbox{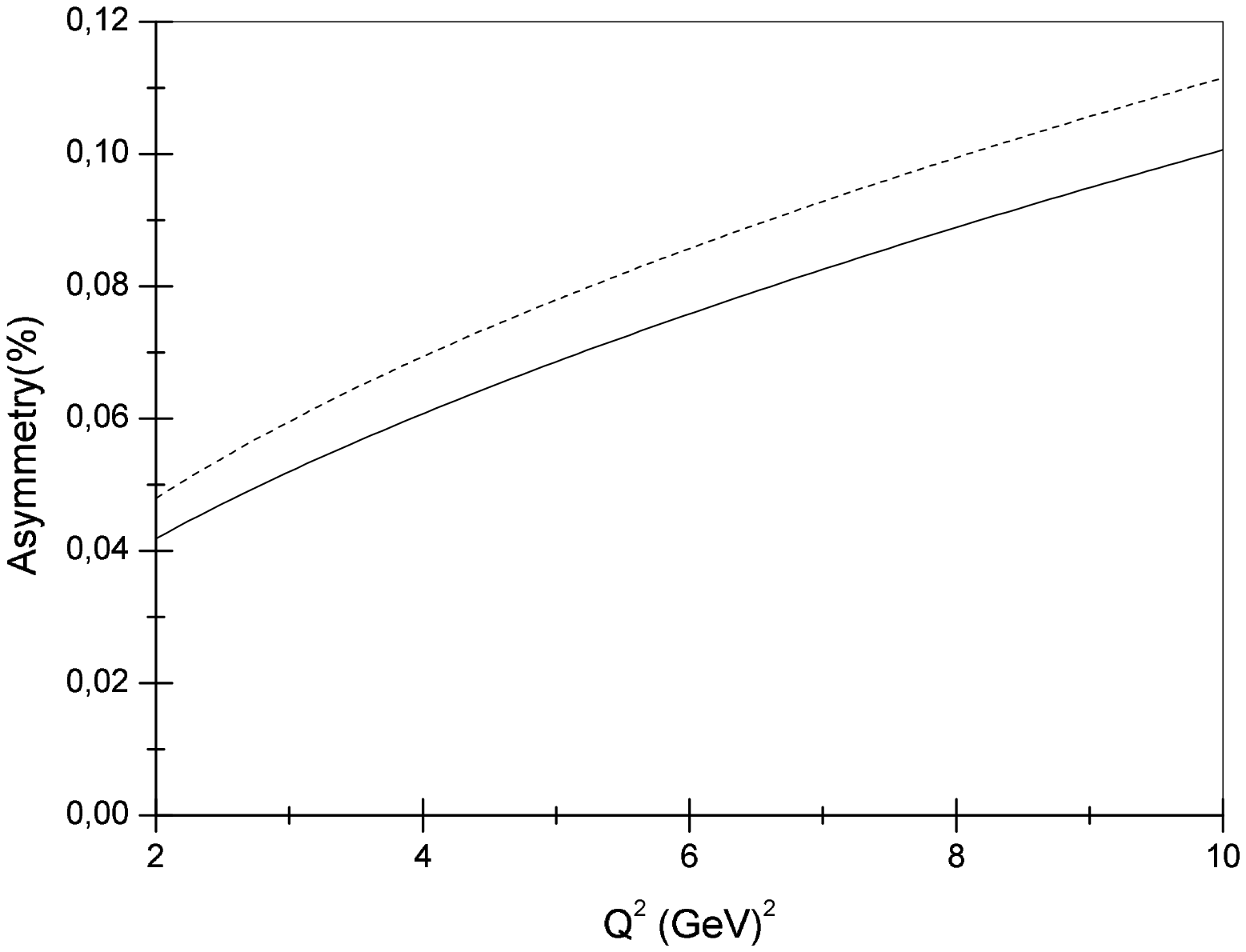}}
\end{minipage}
\begin{minipage}{0.19cm}
\phantom{aa}
\end{minipage}
\begin{minipage}{7.5cm}
\epsfxsize=7.5cm
\centerline{\epsfbox{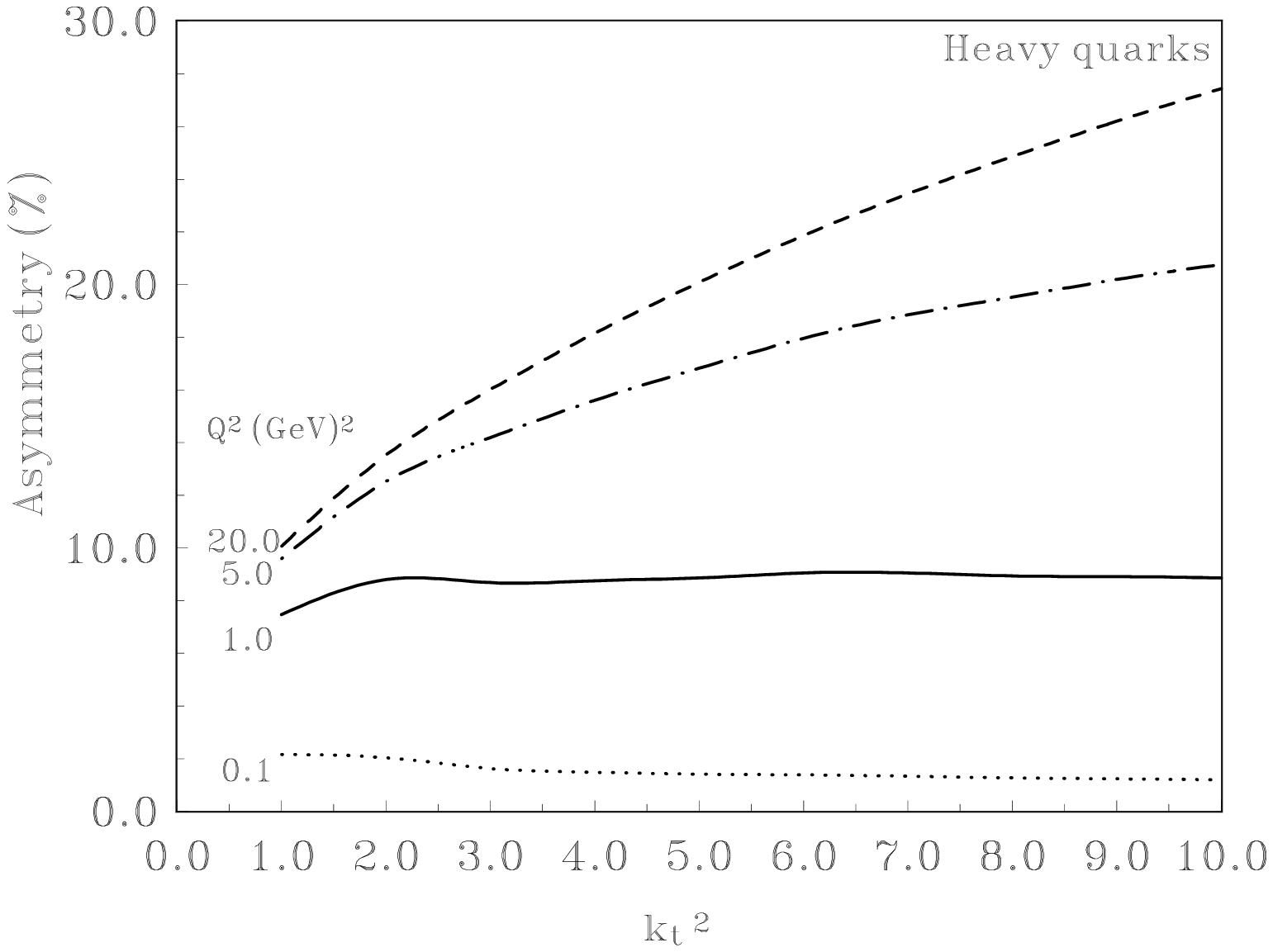}}
\end{minipage}
\\[3mm]
\begin{minipage}{7.5cm}
Fig. 1.~ {The $A_{lT}$ asymmetry for vector meson production
at HERMES ($y$=0.5, $|t|=1 \mbox{GeV}^2$):
solid line -for $J/\Psi$ production; dotted line -for
$\rho$ production.}
\end{minipage}
\begin{minipage}{.15cm}
\phantom{aaa}
\end{minipage}
\begin{minipage}{7.5cm}
Fig. 2.~ {The predicted $Q^2$ dependence of the $A_{lT}$ asymmetry
for the $c \bar c$ production at HERMES for
 $x_P$=0.1, $y$=0.5, $|t|=1 \mbox{GeV}^2$}.
\end{minipage}
\\[4mm]

A similar analysis has been done for the case of $Q \bar Q$ production \cite{gol_j}.
It was found that the asymmetry is proportional, as for vector meson production (\ref{asy}),
to the ratio of
$\tilde K/\tilde B$ functions determined in (\ref{bt}) but with the other scale
$\bar Q^2=m_q^2+\vec k_\perp^2$ ($k_\perp$ is a transverse part of the quark momentum)
.
Now we find two different terms in $d \sigma(-)$:
\begin{equation}\label{ds_qq}
d \sigma^{Q \bar Q}(-)=\left( \vec k_\perp \vec S_\perp \right) d \sigma_k^{Q \bar Q}(-)+
x_P \left( \vec Q_\perp \vec S_\perp \right) d \sigma_k^{Q \bar Q}(-)
\end{equation}
The first term cannot appear for vector meson production. Really in this case, we must
integrate the amplitude over the transverse quark momentum with the meson wave function.
As a result, the term $\propto  (\vec k_\perp \vec S_\perp) $
 disappears, and only the second term which has $x_P$ smallness contributes.
The situation is opposite for $Q \bar Q$ production.
If we can distinguish quark and antiquark jets, with opposite values
of $k_\perp$, the term $\propto \left( \vec k_\perp \vec S_\perp \right)$ contributes.
This term is large and does not have $x_P$ smallness as the second one
in (\ref{ds_qq}).  It gives a predominant contribution to $d \sigma^{Q \bar Q}(-)$.
Note that the $A_{lT}$ asymmetry is proportional to the scalar production of
the proton spin vector, and the jet momentum $A_{lT} \propto
(\vec k_{\perp} \vec S_{\perp}) \propto \cos (\phi_{Jet})$, and the
asymmetry integrated over the azimuthal jet angle $\phi_{Jet}$ is
zero. We have calculated the $A_{lT}$ asymmetry for the case when
the proton spin vector is perpendicular to the lepton scattering
plane and the jet momentum is parallel to this spin vector. The
predicted asymmetry is large, about 10\%, and shown in Fig. 2.

In the present report, the polarized
diffractive hadron leptoproduction at high energies has been
studied within the two-gluon exchange model with the spin-dependent
$gg$-proton coupling (\ref{ver}). This
means that  our results should be applicable for reactions
which include heavy quarks. For processes with light quarks, our
predictions should be valid in high energy range or small $x$
region ($x \le 0.05$ e.g) where the contribution of quark SPD should be small.
The $A_{lT}$ is proportional to the ratio of $\tilde K/\tilde B$
structure functions and generally can be used to get information on the
transverse distribution ${\cal K}^g_{x_P}(x_P,t)$ from experiment.  There are some
difficulties here. The asymmetry for vector meson production is expected
to be quite small $A_{lT}< 0.1 \%$. Similar asymmetry for
$Q \bar Q$ production is predicted to be about 10\%. It is an excellent object to study
transverse effects in the proton coupling with gluons.
However, the experimental study of this asymmetry is not so simple.
To find nonzero asymmetry
in this case, it is necessary to distinguish quark and antiquark jets
and to have possibility to study azimuthal asymmetry structure. This is important,
because cross sections integrated over $d \phi_{Jet}$ are equal to zero.

It is shown that the gluon SPD ${\cal F}^g_{x_P}(x_P)$ and
distribution ${\cal K}^g_{x_P}(x_P)$ at
 small $x_P$ can be studied from the
double spin asymmetry in the hadron leptoproduction. The
contributions of the quark SPDs are non-negligible for $x$ of
about 0.1.
Thus, in the case of the $\rho$ production, the quark SPD
might be studied in addition to the gluon one in the HERMES  experiments.
We conclude that important information on the spin--dependent SPD
at small $x$ can be obtained from the asymmetries in the
diffractive hadron leptoproduction for longitudinally polarized
lepton and transversely polarized hadron targets.

\end{document}